\documentclass[lettersize,journal]{IEEEtran}
\IEEEoverridecommandlockouts
\usepackage{amsfonts}
\usepackage{bbding}
\usepackage{pifont} 
\usepackage{array}

\usepackage{tikz, pgfplots}
 \pgfplotsset{compat=1.18}

\usepackage{textcomp}
\usepackage{stfloats}
\usepackage{url}
\usepackage{verbatim}
\usepackage{graphicx}
\usepackage{xcolor}
\usepackage{subfigure}
\usepackage{algorithm}
\usepackage{algpseudocode}
\usepackage{amsmath}
\usepackage{amssymb}
\usepackage{hyperref}
\usepackage{overpic, scalerel}
\usepackage{bm}
\usepackage{epstopdf}
\hypersetup{
    colorlinks=true,     
    linkcolor=blue,      
    citecolor=blue,      
    urlcolor=blue,       
    pdfborder={0 0 0}    
}
\usepackage{orcidlink}

\usepackage{soul}

\usepackage{cite}

\title{Interference Mitigation in STAR-RIS-Aided Multi-User Networks with Statistical CSI}
\author{Abuzar B. M. Adam$^{_{\orcidlink{0000-0002-9231-9734}}}$,~\IEEEmembership{Member,~IEEE},  Mohammed A. M. Elhassan$^{_{\orcidlink{0000-0003-3522-3055}}}$,  Elhadj Moustapha Diallo$^{_{\orcidlink{0009-0000-4860-6253}}}$, Mohamed Amine Ouamri$^{_{\orcidlink{0000-0002-1922-5483}}}$,~\IEEEmembership{Member,~IEEE}
\thanks{A. B. M. Adam is with the Interdisciplinary Centre for Security, Reliability and Trust (SnT), University of Luxembourg, 1855 Luxembourg City, Luxembourg (e-mail: abuzar.babikir@uni.lu).}
\thanks{Mohammed A. M. Elhassan is with School of Computer Science and Technology, Zhejiang Normal University (e-mail: Mohammedac29@zjnu.edu.cn).}
\thanks{E. M. Diallo is with the School of Communications and Information Engineering, Chongqing University of Posts and Telecommunications, Chongqing, China.}
\thanks{M. Ouamri is with Depratement of Network and Télécommunications, IUT de Villetaneuse, L2TI Laboratory, Sorbonne Paris North University, 93400, France (email:ouamrimouhamedamine@gmail.com).}
 }

\begin{document}
\maketitle

\begin{abstract}
In this paper, we investigate real-time interference mitigation in multiuser wireless networks assisted by simultaneously transmitting and reflecting reconfigurable intelligent surfaces (STAR-RISs). Unlike conventional methods that rely on instantaneous channel state information (CSI), we consider a practical scenario where only statistical CSI is available, and the STAR-RIS phase shifts are impaired by random phase errors modeled via the Von Mises distribution. To tackle the resulting nonconvex optimization problem induced by unit-modulus constraints and stochastic interference, we derive a closed-form approximation of the effective channel matrix using statistical expectations. We then reformulate the interference minimization problem as an unconstrained optimization over a Riemannian manifold and propose a conjugate gradient algorithm tailored to the complex circle manifold. The proposed solution enables efficient real-time computation of optimal phase shifts while accounting for hardware imperfections and limited CSI. Simulation results confirm that our method significantly suppresses inter-user interference and achieves superior SINR performance and convergence speed compared to conventional baselines.
\end{abstract}

\section{Introduction}

As sixth-generation (6G) wireless technology continues to evolve, the demand for high-capacity wireless data services is growing rapidly \cite{11023218,adam2024ad,10287263}. To accommodate this surge in traffic requirements, a wide range of advanced techniques are being explored for future 6G communication systems \cite{9805661}. Among these, reconfigurable intelligent surfaces (RISs) have emerged as a transformative technology for beyond-5G and 6G wireless networks, offering unprecedented control over the radio environment through programmable manipulation of electromagnetic waves \cite{10829753}.

Unlike conventional wireless networks where the propagation environment is considered fixed and uncontrollable, RISs, also referred to as intelligent reflecting surfaces, introduce a new communication paradigm in which the wireless environment itself becomes programmable. By dynamically adjusting the phase shifts of incident signals, RISs can shape the propagation characteristics to create favorable transmission conditions.

A particularly promising variant of this technology is the simultaneously transmitting and reflecting RIS (STAR-RIS), which extends the reflection-only paradigm by enabling energy to be both transmitted and reflected toward different user groups \cite{11023218}. This dual-functionality greatly enhances spectral and energy efficiency, making STAR-RIS a compelling enabler for full-space coverage and multiuser connectivity in next-generation networks \cite{10802977}.

Despite its potential, practical deployment of STAR-RIS systems faces several challenges. First, the acquisition of instantaneous channel state information (CSI) is particularly difficult due to the passive nature of the surface and the large number of elements \cite{10494543}. Second, phase shifts at the STAR-RIS are often impaired by hardware imperfections, such as phase noise, which can degrade the overall performance \cite{10841966}. Third, the interference among simultaneously served users becomes non-negligible in dense networks, requiring sophisticated interference mitigation strategies. These challenges call for robust optimization frameworks that can operate under partial CSI and hardware constraints.

In this paper, we focus on real-time interference mitigation in STAR-RIS-aided multiuser systems by leveraging statistical CSI and geometric optimization tools. Specifically, we model the STAR-RIS phase noise using a Von Mises distribution and derive a closed-form approximation of the effective channel matrix in the presence of random phase errors. The resulting optimization problem, which aims to minimize the expected inter-user interference under unit-modulus constraints, is inherently nonconvex. To address this, we propose a Riemannian manifold optimization framework that enables unconstrained gradient-based updates on the complex circle manifold. Simulation results validate the efficacy of the proposed method, demonstrating superior interference suppression and fast convergence compared to conventional baselines.

The integration of STAR-RIS technology into wireless networks promises significant enhancements in signal coverage and spectral efficiency. However, most existing works rely heavily on instantaneous CSI and assume ideal hardware conditions, assumptions that are difficult to meet in practical systems with passive surfaces and limited feedback capabilities. Moreover, traditional optimization techniques such as alternating optimization and successive convex approximation (SCA) often incur high computational complexity and lack robustness under CSI uncertainty.

These gaps motivate the need for a new class of optimization frameworks that can (i) operate using only statistical CSI, (ii) account for realistic impairments such as phase noise, and (iii) achieve real-time interference mitigation in multiuser environments. Our work addresses these needs by combining a statistical CSI model with Riemannian optimization on the complex circle manifold, thereby enabling efficient and practical STAR-RIS configuration in next-generation wireless networks.
The main contributions of this paper are summarized as follows:

\begin{itemize}
    \item We formulate a STAR-RIS-aided multiuser communication model under statistical CSI and phase noise, reflecting practical scenarios where instantaneous CSI acquisition is infeasible due to the passive nature of the RIS.
    
    \item The STAR-RIS phase shifts are modeled as being corrupted by hardware-induced phase noise, captured analytically using the Von Mises distribution to characterize random phase deviations.

    \item A closed-form approximation of the effective channel matrix is derived based on statistical expectations, which facilitates tractable reformulation of the interference mitigation problem without relying on instantaneous CSI.

    \item The resulting nonconvex optimization problem is reformulated as an unconstrained problem on a complex circle Riemannian manifold, enabling the application of efficient geometric optimization techniques.

    \item We propose a Riemannian conjugate gradient algorithm that operates on the manifold’s tangent space and incorporates appropriate projection and retraction operations for efficient update of STAR-RIS phase configurations.

    \item Extensive simulation results demonstrate that the proposed approach achieves effective inter-user interference suppression and faster convergence compared to conventional baselines under practical CSI and hardware constraints.
\end{itemize}

\section{System Model and Problem Formulation}
\begin{figure}[!ht] 
\centerline{\includegraphics[width=3.5 in]{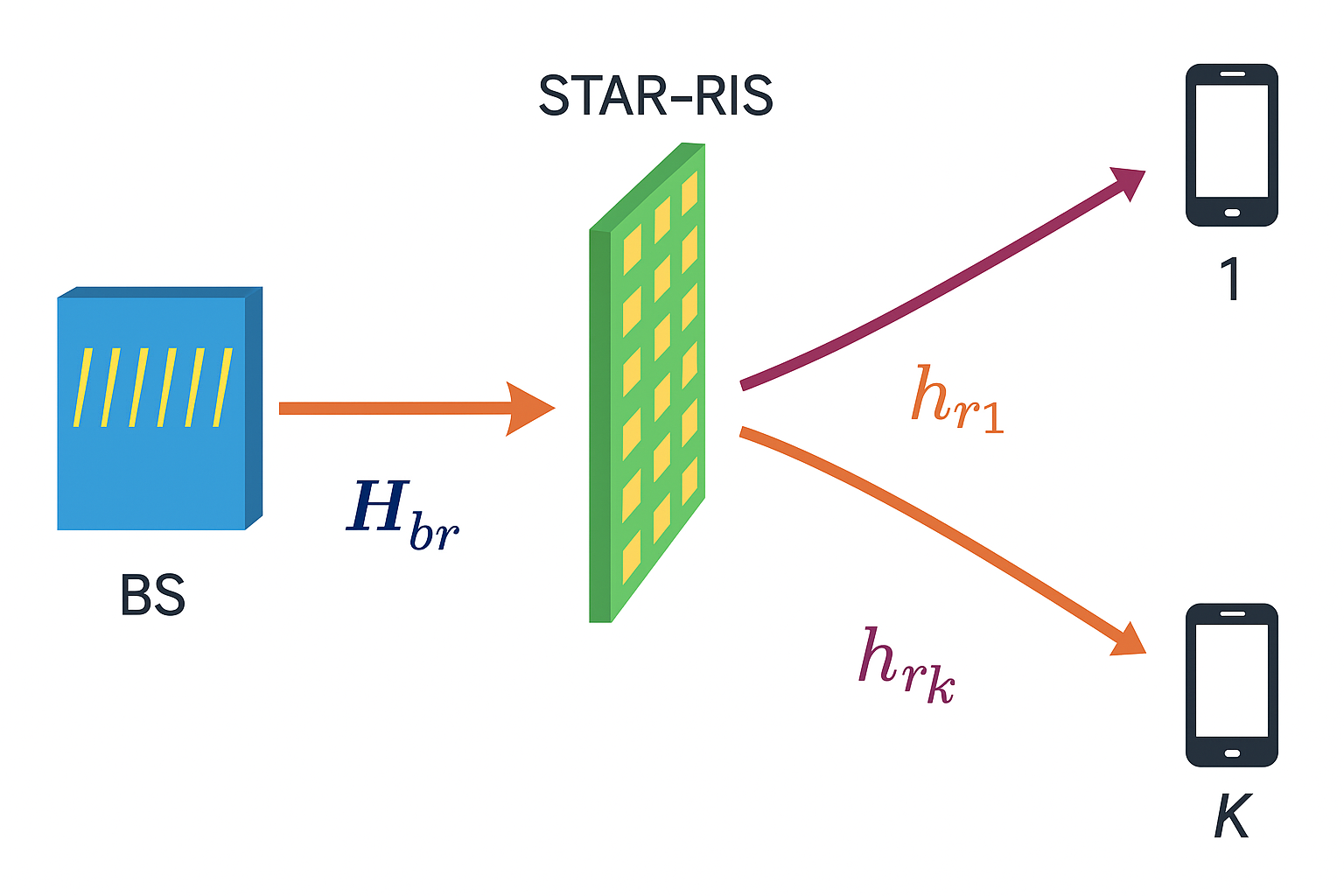}} 
\caption{Illustration of STAR-RIS-aided multiuser network.} 
\label{sys} \end{figure}

In this work, we consider a STAR-RIS-assisted multiuser network where a BS  is communicating with $K$ single antenna users as illustrated in \ref{sys}.  BS $b$ equipped with $M$ antennas forming uniform linear array (ULA). The STAR-RIS is equipped with $N$ reflecting elements forming a uniform planar array (UPA). Based on their location to the STAR-RIS, the users are classified as either transmitting users or receiving users.

\subsection{Channel and Signal Model}
The channel coefficients between the BS $b$ and user $k$, BS and RIS, and RIS and the user $k$ are given as follows
\begin{equation}\label{eqn1}
{{\bf{h}}_{bk}} = \sqrt {{\alpha _{bk}}} {{\bf{g}}_{bk}} \in {\mathbb{C}^{M \times 1}}
\end{equation}
\begin{equation}\label{eqn2}
{{\bf{H}}_{br}} = \sqrt {{\alpha _{br}}} {{\bf{G}}_{br}} \in {\mathbb{C}^{M \times N}}
\end{equation}
\begin{equation}\label{eqn3}
{\bf{h}}_{rk}^i = \sqrt {{\alpha _{rk}}} {\bf{g}}_{rk}^i \in {\mathbb{C}^{N \times 1}}
\end{equation}
where ${\alpha _l},l \in \left\{ {bk,br,rk} \right\}$ denotes the large-scale fading coefficient and ${{\bf{g}}_l},l \in \left\{ {bk,br,rk} \right\}$ denotes the fast-fading channel coefficient. $i \in \left\{ {tr,re} \right\}$ denotes the transmitting or receiving mode. Assuming the channels are following Rician fading, the fast-fading channels can be expressed as follows
\begin{equation}\label{eqn4}
{{\bf{g}}_{bk}} = \sqrt {\frac{{{\beta _{bk}}}}{{{\beta _{bk}} + 1}}} {{\bf{\bar g}}_{bk}} + \sqrt {\frac{1}{{{\beta _{bk}} + 1}}} {{\bf{\tilde g}}_{bk}},
\end{equation}
\begin{equation}\label{eqn5}
{{\bf{G}}_{br}} = \sqrt {\frac{{{\beta _{br}}}}{{{\beta _{br}} + 1}}} {{\bf{\bar G}}_{br}} + \sqrt {\frac{1}{{{\beta _{br}} + 1}}} {{\bf{\tilde g}}_{br}},
\end{equation}
\begin{equation}\label{eqn6}
{\bf{g}}_{rk}^i = \sqrt {\frac{{{\beta _{rk}}}}{{{\beta _{rk}} + 1}}} {{\bf{\bar g}}_{rk}} + \sqrt {\frac{1}{{{\beta _{rk}} + 1}}} {{\bf{\tilde g}}_{rk}},
\end{equation}
where $\beta_l$ is the Rician factor and ${{\bf{\bar g}}_l}$ and  ${{\bf{\tilde g}}_l}$ are the LoS and NLoS components follow circularly symmetric complex Gaussian distribution ${\mathcal{C}}{\mathcal{N}} \sim \left( {0,1} \right)$. ${{\bf{\bar g}}_l}$ for different links are given as follows
\begin{equation}\label{eqn7}
\begin{aligned}
{{\bf{\bar g}}_{bk}} = \Big[\, &1,\; \ldots,\; 
\exp\left( j\frac{2\pi}{\lambda}a \sin{\phi_{bk}} \right),\; \ldots, \\
&\exp\left( j\frac{2\pi}{\lambda}a(M - 1)\sin{\phi_{bk}} \right) 
\,\Big]^T
\end{aligned}
\end{equation}

\begin{equation}\label{eqn8}
\begin{aligned}
\bar{\mathbf{g}}_{rk} = \Big[\, &1,\; \ldots,\; 
\exp\left( j\frac{2\pi}{\lambda}d \left( 
\sin{\phi_{rk}} \sin{\psi_{rk}} \right.\right. \\
&\left.\left. +\; \cos{\phi_{rk}} \sin{\psi_{rk}} \right) \right),\; \ldots\, \Big]^T
\end{aligned}
\end{equation}

\begin{equation}\label{eqn9}
{{\bf{\bar G}}_{br}} = \Upsilon _N^T \otimes {\Upsilon _M},
\end{equation}
where
\begin{equation*}
    \begin{aligned}
        {\Upsilon _N} &= \left[ {1,...,\exp \left( {j\frac{{2\pi d}}{\lambda }x\sin \phi \cos \varphi } \right)} \right]\\ 
        &\otimes \left[ {1,...,\exp \left( {j\frac{{2\pi d}}{\lambda }z\cos \phi } \right)} \right],
    \end{aligned}
\end{equation*}
\begin{equation*}
{\Upsilon _M} = \left[ {1,\exp \left( {j\frac{{2\pi }}{\lambda }a\cos \theta } \right),...,\exp \left( {j\frac{{2\pi }}{\lambda }a\left( {M - 1} \right)\cos \theta } \right)} \right],
\end{equation*}
where $a$ is the antenna separation and $d = {\textstyle{\lambda  \over 2}}$ is the RIS elements separation. The passive beamforming matrix of the RIS is given as ${\Theta ^i} = \text{diag}\left[ {{e^{j\theta _1^i}},{e^{j\theta _2^i}},...,{e^{j\theta _N^i}}} \right],i \in \left\{ {tr,re} \right\}$ and the phase noise matrix is given as ${\Phi ^i} = \left[ {{e^{j\Delta \theta _1^i}},{e^{j\Delta \theta _2^i}},...,{e^{j\Delta \theta _n^i}},...,{e^{j\Delta \theta _N^i}}} \right],i \in \left\{ {tr,re} \right\}$ with ${e^{j\Delta \theta _n^i}}$ is a random phase error follows Von Mises distribution with zero mean and concentration  ${\varepsilon _{\Delta \theta }}$ where $\Delta \theta _n^i$ has a characteristic function $\chi  = {\textstyle{{{I_1}\left( {{\varepsilon _{\Delta \theta }}} \right)} \over {{I_0}\left( {{\varepsilon _{\Delta \theta }}} \right)}}}$ with  ${I_p}\left( {{\varepsilon _{\Delta \theta }}} \right)$ denotes the modified first type Bessel function with the order $p$.
The received signal at the user $k$ is given as
\begin{equation}\label{eqn10}
\begin{aligned}
{y_k} &= \left( {{\bf{h}}_{bk}^H+{\bf{H}}_{br}^{i,H}{\Theta ^i}{\Phi ^i}{\bf{h}}_{rk}^i} \right)\sum\limits_{k = 1}^K {{{\bf{w}}_k}{{\bf{s}}_k}}  + {\eta _k}\\
 &= \underbrace {\left( {{\bf{h}}_{bk}^H+{\bf{H}}_{br}^{i,H}{\Theta ^i}{\Phi ^i}{\bf{h}}_{rk}^i} \right){{\bf{w}}_k}{{\bf{s}}_k}}_{\text{Desired signal}}\\ 
 &+ \underbrace {\left( {{\bf{h}}_{bk}^H+{\bf{H}}_{br}^{i,H}{\Theta ^i}{\Phi ^i}{\bf{h}}_{rk}^i} \right)\sum\limits_{j = 1,j \ne k}^K {{{\bf{w}}_j}{{\rm{s}}_j}} }_{\text{Inter-user interference}}
+ \underbrace {{\eta _k}}_{\text{Noise}},
\end{aligned}
\end{equation}
where $\mathbb{E}\left[ {{{\left| {{{\bf{s}}_k}} \right|}^2}} \right] = 1$  and ${\eta _k} \sim {\cal C}{\cal N}\left( {0,{\sigma ^2}} \right)$ is the additive while Gaussian noise (AWGN).
\subsection{Problem Formulation}
In the literature, the effective channel matrix is used to perform the interference mitigation \cite{9090356}. Thus, the received signal can be expressed as
\begin{equation}\label{eqn11}
\begin{array}{l}
{\bf{y}} = \left( {{{\bf{H}}_{bK}} + {{\bf{H}}_{rK}}{{\bf{\Theta }}^i}{\Phi ^i}{{\bf{H}}_{br}}} \right){\bf{S}} + {\bf{{\bf E}}}\\
 = {{\bf{G}}_{\rm eff}}{\bf{S}} + {{{\bf E}}}
\end{array}
\end{equation}
where ${\bf{y}} = \left[ {{y_1},{y_2},...,{y_k}} \right]$, ${{\bf{H}}_{bK}} \in {\mathbb{C}^{M \times K}}$ contains all the channel coefficients between the BS $b$ and the users,  ${{\bf{H}}_{rK}} \in {\mathbb{C}^{N \times K}}$ contains all the channel coefficients between the RIS and the users, ${{\bf{H}}_{br}} \in {\mathbb{C}^{M \times N}}$ comprises all the channel coefficients between $b$ and the RIS. ${\bf{S}} = \left[ {{{\bf{s}}_1},{{\bf{s}}_2},...,{{\bf{s}}_K}} \right]$ and ${\bf{{\bf E}}} = \left[ {{\eta _1},{\eta _2},...,{\eta _K}} \right]$. ${{\bf{G}}_{\rm eff}}$ constitutes the effective channel matrix. In this context, the signal-to-interference-plus-noise ratio (SINR) of the user $k$ is given as
\begin{equation}\label{eqn12}
{\gamma _k} = \frac{{{{\left| {{{\bf{G}}_{{\rm eff},k,k}}} \right|}^2}}}{{\sum\limits_{j = 1,j \ne k}^K {{{\left| {{{\bf{G}}_{{\rm eff},k,j}}} \right|}^2}}  + {\sigma ^2}}}
\end{equation}

However, the exact effective channel matrix is difficult obtain especially in presence phase shift error. Moreover, from practical point of view, acquiring the instantaneous CSI might be unattainable due to the passive nature of the RIS \cite{10075533,adam2023}. Therefore, we resort to the estimation of the effective channel matrix. The problem can be formulated as follows
\begin{subequations}\label{eqn13:main}
\begin{align}
&\mathop {\min }\limits_{\theta} && \mathbb{E}\left\{ \sum\limits_{j = 1,j \ne k}^K {{{\left| {{{\bf{G}}_{{\rm eff},k,j}}} \right|}^2}} \right\}   &   & \tag{\ref{eqn13:main}} \\
& \text{s.t.}&& \left| {{e^{j\left( {\theta _n^i + \Delta \theta _n^i} \right)}}} \right| = 1,\forall n = 1,2,..,N,i \in \left\{ {\rm tr,re} \right\},\notag
\end{align}
\end{subequations}
Problem \eqref{eqn13:main} is nonconvex due to the probabilistic expression of the utility the unit-modulus constraints.

\subsection{Problem Reformulation}
The first step is to obtain a closed-form expression for the utility function in \eqref{eqn13:main} using the following theorem.

\emph{Theorem 1:} The effective channel can be approximated to the following closed-form expression
\begin{equation}\label{eqn14}
\begin{aligned}
\mathbb{E}\left\{ {{{\bf{G}}_{{\rm eff},k,j}}} \right\} &= \frac{{\sqrt {{\alpha _{bk}}} \sqrt {{\beta _{bk}}} {\Lambda _m} + {\xi _m}}}{{\sqrt {{\beta _{bk}} + 1} }}\\
 &+ \sqrt {{\alpha _{br}}} \sqrt {{\alpha _{rk}}} \frac{{\sqrt {{\beta _{br}}} \sqrt {{\beta _{rk}}} {\Lambda _{mn}} + \sqrt {{\beta _{br}}}  + \sqrt {{\beta _{rk}}}  + 1}}{{\sqrt {\left( {{\beta _{br}} + 1} \right)\left( {{\beta _{rk}} + 1} \right)} }},
\end{aligned}
\end{equation}
\emph{Proof:} Please refer to Appendix I

Problem \eqref{eqn13:main} can be reformulated as follows
\begin{subequations}\label{eqn14:main}
\begin{align}
&\mathop {\min }\limits_{\boldsymbol\upsilon} && \sum\limits_{k = 1}^K {{{\left\| \begin{array}{l}
\frac{{\sqrt {{\alpha _{bk}}} \sqrt {{\beta _{bk}}} {\Lambda _m} + {\xi _m}}}{{\sqrt {{\beta _{bk}} + 1} }}\\
 + \sqrt {{\alpha _{br}}} \sqrt {{\alpha _{rk}}} \frac{{\sqrt {{\beta _{br}}} \sqrt {{\beta _{rk}}} {\Lambda _{mn}} + \sqrt {{\beta _{br}}}  + \sqrt {{\beta _{rk}}}  + 1}}{{\sqrt {\left( {{\beta _{br}} + 1} \right)\left( {{\beta _{rk}} + 1} \right)} }}
\end{array} \right\|}^2}}    &   & \tag{\ref{eqn14:main}} \\
& \text{s.t.}&& \left| {\upsilon _n^i} \right| = 1,\forall n = 1,2,..,N,i \in \left\{ {\rm tr,re} \right\},\label{eqn14:a}
\end{align}
\end{subequations}
where $\upsilon _n^i = {e^{j\left( {\theta _n^i + \Delta \theta _n^i} \right)}}$. Problem \eqref{eqn14:main} still nonconvex due to the unit-modulus constraints and cannot be solved by the conventional methods such as SCA. In the next sections, we propose a solution for problem \eqref{eqn14:main}.

\section{Riemannian Manifold Optimization Framework}
In this section, we aim at circumventing the main issue of the unit-modulus constraints and the transformation of problem \eqref{eqn14:main} into unconstrained optimization problem.

According to \cite{9405423}, the unit-modulus constraints \eqref{eqn14:a} can be considered as a restriction of the solution to lie on the surface of a smooth Riemannian manifold embedded in $\mathbb{C}^N$. Since there are no other constraints, the problem can be reformulated as unconstrained problem on the surface of complex circle manifold where each element $\upsilon _n^i$ lives on surface of this complex circle.
The unconstrained problem is given as follows~\cite{10925865,Li2025DFP}
\begin{equation}\label{eqn15}
\mathop {\min }\limits_{\upsilon  \in {\mathcal{R}}} f\left( {{\boldsymbol \upsilon }} \right),
\end{equation}
where $\mathcal{R}$ is a Riemannian manifold and the objective function $f:{\mathcal{R}} \to \mathbb{R}$ is given as
\begin{equation}\label{eqn16}
f\left( {{\boldsymbol\upsilon }} \right) = \sum\limits_{k = 1}^K {{{\left\| \begin{array}{l}
\frac{{\sqrt {{\alpha _{bk}}} \sqrt {{\beta _{bk}}} {\Lambda _m} + {\xi _m}}}{{\sqrt {{\beta _{bk}} + 1} }}\\
 + \sqrt {{\alpha _{br}}} \sqrt {{\alpha _{rk}}} \frac{{\sqrt {{\beta _{br}}} \sqrt {{\beta _{rk}}} {\Lambda _{mn}} + \sqrt {{\beta _{br}}}  + \sqrt {{\beta _{rk}}}  + 1}}{{\sqrt {\left( {{\beta _{br}} + 1} \right)\left( {{\beta _{rk}} + 1} \right)} }}
\end{array} \right\|}^2}} ,
\end{equation}
Problem \eqref{eqn15} can be solved using gradient descent on Riemannian manifold. In particular, first we calculate the Euclidean gradient. Let ${{\bf{p}}^{\left( t \right)}} \in \mathbb{C}:{{\bf{p}}^{\left( t \right),*}}{{\bf{p}}^{\left( t \right)}} = 1$ denotes point in the tangent space on the complex circle, the space of all tangent vectors passing through ${{\bf{p}}^{\left( t \right)}}$ is given as ${{\cal T}_{{p_n},{\cal R}}} = \left\{ {{\bf{Q}} \in {\mathbb{C}^N}:\Re \left( {{\bf{Q}} \odot {{\bf{p}}^{\left( t \right),*}} = {{\mathbf{0}}_N}} \right)} \right\}$, the Euclidean gradient of $f\left( {{{\bf{p}}^{\left( t \right)}}} \right)$ is given as \eqref{eqn17} on top of next page,
\begin{figure*}
    \begin{equation}\label{eqn17}
\nabla f\left( {{{\bf{p}}^{\left( t \right)}}} \right) = \sum\limits_{k = 1}^K {2{\bf{\bar D}}\left[ 
\frac{{\sqrt {{\alpha _{bk}}} \sqrt {{\beta _{bk}}} {\Lambda _m} + {\xi _m}}}{{\sqrt {{\beta _{bk}} + 1} }}
 + \sqrt {{\alpha _{br}}} \sqrt {{\alpha _{rk}}} \frac{{\sqrt {{\beta _{br}}} \sqrt {{\beta _{rk}}} {\Lambda _{mn}} + \sqrt {{\beta _{br}}}  + \sqrt {{\beta _{rk}}}  + 1}}{{\sqrt {\left( {{\beta _{br}} + 1} \right)\left( {{\beta _{rk}} + 1} \right)} }}
 \right]} ,
\end{equation}
\hrule
\end{figure*}
where the detailed definition of ${\bf{\bar D}}$ is given as in \eqref{details} on top of the next page.
\begin{figure*}
\begin{equation}\label{details}
\begin{array}{l}
{\bf{\bar D}} = \frac{{\sqrt {{\alpha _{br}}} \sqrt {{\alpha _{rk}}} \sqrt {{\beta _{br}}} \sqrt {{\beta _{rk}}} {\bf{\tilde N}}}}{{\sqrt {\left( {{\beta _{br}} + 1} \right)\left( {{\beta _{rk}} + 1} \right)} }},\\
{\bf{\tilde N}} =  - 2{\chi ^2}\frac{{{{\tilde {\rm N}}_1}{{\tilde {\rm N}}_2} - {{\tilde {\rm N}}_3}{{\tilde {\rm N}}_4}}}{{\tilde {\rm N}_1^2}},\\
{{\tilde {\rm N}}_1} = \sum\limits_{m,n = 1}^{MN} {\cos \left( {\theta _n^i - \theta _m^i + \left( {\frac{{2\pi }}{\lambda }a\left( {\left( {{x_n} - {x_m}} \right){{\cal A}_{br}} + \left( {{z_n} - {z_m}} \right){{\cal B}_{br}}} \right)} \right)} \right)} ,\\
{{\tilde {\rm N}}_2} = \sum\limits_{1 \le q < n \le N}^N {\exp \left( {j\frac{{2\pi }}{\lambda }d\left( {\left( {{x_n} - {x_q}} \right){{\cal A}_{rk}} + \left( {{z_n} - {z_q}} \right){{\cal B}_{rk}}} \right)} \right)} ,\\
{{\tilde {\rm N}}_3} = \sum\limits_{1 \le q < n \le N}^N {\cos \left( {\theta _n^i - \theta _q^i + \left( {\frac{{2\pi }}{\lambda }d\left( {\left( {{x_n} - {x_q}} \right){{\cal A}_{rk}} + \left( {{z_n} - {z_q}} \right){{\cal B}_{rk}}} \right)} \right)} \right)} ,\\
{{\tilde {\rm N}}_4} = \sum\limits_{m,n = 1}^{MN} {\exp \left( {j\frac{{2\pi }}{\lambda }a\left( {\left( {{x_n} - {x_m}} \right){{\cal A}_{br}} + \left( {{z_n} - {z_m}} \right){{\cal B}_{br}}} \right)} \right)}
\end{array}
\end{equation}
\hrule
\end{figure*}

The Riemannian gradient of $f\left( {{{\bf{p}}^{\left( t \right)}}} \right)$ on the complex circle manifold is given as a projection of Euclidean gradient from the ambient Euclidean space onto the tangent space ${{\cal T}_{{{\bf{p}}^{\left( t \right)}},{\cal R}}}$ at the point ${{\bf{p}}^{\left( t \right)}}$ on the complex circle which can be expressed as
\begin{equation}\label{eqn18}
{\nabla _{\cal R}}f\left( {{{\bf{p}}^{\left( t \right)}}} \right) = \nabla f\left( {{{\bf{p}}^{\left( t \right)}}} \right) - \Re \left\{ {\nabla f\left( {{{\bf{p}}^{\left( t \right)}}} \right) \odot {{\bf{p}}^{\left( t \right),*}}} \right\} \odot {{\bf{p}}^{\left( t \right)}},
\end{equation}
\begin{algorithm}[H]
\label{alg}
\caption {Riemannian Manifold Optimization for Interference Mitigation in STAR-RIS Networks}
\begin{algorithmic}[1]
\renewcommand{\algorithmicrequire}{\textbf{Initialization:}}
\Require ${{\bf{p}}^{\left( 0 \right)}}, t = 0$ $\epsilon = 0.001$
\While{$\left| {{\nabla _{\cal R}}f\left( {{{\bf{p}}^{\left( t \right)}}} \right)} \right| <  \epsilon$}
\State Compute $\nabla f\left( {{{\bf{p}}^{\left( t \right)}}} \right)$ using \eqref{eqn17};
\State  Compute ${{\nabla _{\cal R}}f\left( {{{\bf{p}}^{\left( t \right)}}} \right)}$ using \eqref{eqn18};
\State  Compute $\beta _{{\rm PR} - {\cal R}}^{\left( {t + 1} \right)}$ using \eqref{eqn20};
\State  Compute ${\mathfrak{K}_{{{\bf{p}}^{\left( {t + 1} \right)}}}}$ using \eqref{eqn19};
\State  Update ${{\bf{p}}^{\left( {t + 1} \right)}}$ via \eqref{eqn21};
\State  $t = t + 1$;
\EndWhile
\State  ${{\boldsymbol\upsilon }} = {{\bf{p}}^{\left( t \right)}}$
\end{algorithmic}
\end{algorithm}
To update the direction of the gradient ${\mathfrak{K}_{{{\bf{p}}^{\left( {t + 1} \right)}}}}$, we adopt Polak-Ribi\`{e}re conjugate gradient method as follows
\begin{equation}\label{eqn19}
{\mathfrak{K}_{{{\bf{p}}^{\left( {t + 1} \right)}}}} =  - {\nabla _{\cal R}}f\left( {{{\bf{p}}^{\left( {t + 1} \right)}}} \right) + \beta _{{\rm PR} - {\cal R}}^{\left( {t + 1} \right)}{{\cal T}_{{{\bf{p}}^{\left( t \right)}} \mapsto {{\bf{p}}^{\left( {t + 1} \right)}}}}\left( {{\mathfrak{K}_{{{\bf{p}}^{\left( {t + 1} \right)}}}}} \right),
\end{equation}
where ${{\cal T}_{{{\bf{p}}^{\left( t \right)}} \mapsto {{\bf{p}}^{\left( {t + 1} \right)}}}}\left( {{\mathfrak{K}_{{{\bf{p}}^{\left( {t + 1} \right)}}}}} \right)$ obtained via projection similar to that in \eqref{eqn18} and the manifold generalization of Polak-Ribi\`{e}re conjugate parameter $\beta _{{\rm PR} - {\cal R}}^{\left( {t + 1} \right)}$ is given as follows
\begin{equation}\label{eqn20}
\begin{aligned}
\beta _{{\rm{PR}} - R}^{\left( {t + 1} \right)} &= \frac{{\nabla f\left( {{{\bf{p}}^{\left( {t + 1} \right)}}} \right) - {T_{{{\bf{p}}^{\left( t \right)}} \mapsto {{\bf{p}}^{\left( {t + 1} \right)}}}}\left( {\nabla f\left( {{{\bf{p}}^{\left( t \right)}}} \right)} \right)}}{{{{\left\| {\nabla f\left( {{{\bf{p}}^{\left( t \right)}}} \right)} \right\|}^2}}}\\
 &\qquad{}\times {\left[ {\nabla f\left( {{{\bf{p}}^{\left( {t + 1} \right)}}} \right)} \right]^H},
\end{aligned}
\end{equation}

The final step is to update ${{\bf{p}}^{\left( {t + 1} \right)}}$ which is achieved using the following
\begin{equation}\label{eqn21}
{{\bf{p}}^{\left( {t + 1} \right)}} = \left( {{{\bf{p}}^{\left( t \right)}} + {\delta ^{\left( t \right)}}{\mathfrak{K}_{{{\bf{p}}^{\left( {t + 1} \right)}}}}} \right) \oslash \left| {{{\bf{p}}^{\left( t \right)}} + {\delta ^{\left( t \right)}}{\mathfrak{K}_{{{\bf{p}}^{\left( {t + 1} \right)}}}}} \right|,
\end{equation}
where ${\delta ^{\left( t \right)}}$ is step which is calculated using Armijo backtracking line search algorithm, $\oslash$ indicates Hadamard element-wise division, and $\left|  \cdot  \right|$ indicates element-wise absolute value. Algorithm 1 shows the procedure of solving \eqref{eqn14:main}.

The proposed Riemannian manifold optimization algorithm has a per-iteration complexity of $\mathcal{O}(KMN)$, dominated by gradient evaluations over $K$ users and $N$ RIS elements with $M$ BS antennas. In contrast, the conventional SCA method requires solving a convex program with complexity $\mathcal{O}(N^3)$ per iteration. As a result, the total complexity of the RMO algorithm, $\mathcal{O}(TKMN)$, grows linearly with the system size and is significantly lower than the $\mathcal{O}(L N^3)$ complexity of SCA, making the proposed solution more scalable and suitable for real-time STAR-RIS applications.
\section{Numerical Results}
In this section, we evaluate the performance of the proposed Riemannian manifold optimization algorithm for interference mitigation in STAR-RIS-aided multiuser networks. The simulation considers a downlink communication scenario where a multi-antenna BS serves single-antenna users via a STAR-RIS with simultaneous transmission and reflection functionality. The channels are modeled with Rician fading, and phase noise at the STAR-RIS elements is incorporated using a Von Mises distribution. The objective is to minimize the expected inter-user interference under statistical CSI and hardware impairments. The performance of the proposed algorithm is compared against conventional benchmarks, including SCA-based optimization. All results are averaged over multiple channel realizations to ensure statistical reliability.
The key parameters used in the simulation are summarized in Table~\ref{table:sim_parameters}, unless otherwise state.
\begin{table}[h]
\centering
\caption{Simulation Parameters}
\label{table:sim_parameters}
\begin{tabular}{l|c}
\hline
\textbf{Parameter} & \textbf{Value} \\
\hline
Number of BS antennas ($M$) & 8 \\
Number of users ($K$) & 4 \\
Number of STAR-RIS elements ($N$) & 64 \\
RIS element spacing ($d$) & $\lambda/2$ \\
BS antenna spacing ($a$) & $\lambda/2$ \\
Carrier frequency & 28 GHz \\
Wavelength ($\lambda$) & $c/f_c$ \\
Rician factor ($\beta_{l}$) & 10 dB \\
Large-scale fading ($\alpha_{bk}, \alpha_{br}, \alpha_{rk}$) & 0 to –20 dB (distance-based) \\
Noise power spectral density & –174 dBm/Hz \\
Bandwidth & 10 MHz \\
Modulation & QPSK \\
Phase noise model & Von Mises (mean = 0, $\varepsilon_{\Delta\theta} = 10$) \\
Convergence tolerance ($\epsilon$) & $10^{-3}$ \\
Max iterations (Proposed) & 200 \\
Max iterations (SCA) & 200 \\
\hline
\end{tabular}
\end{table}

\begin{figure}[!ht] 
\centerline{\includegraphics[width=3.5 in]{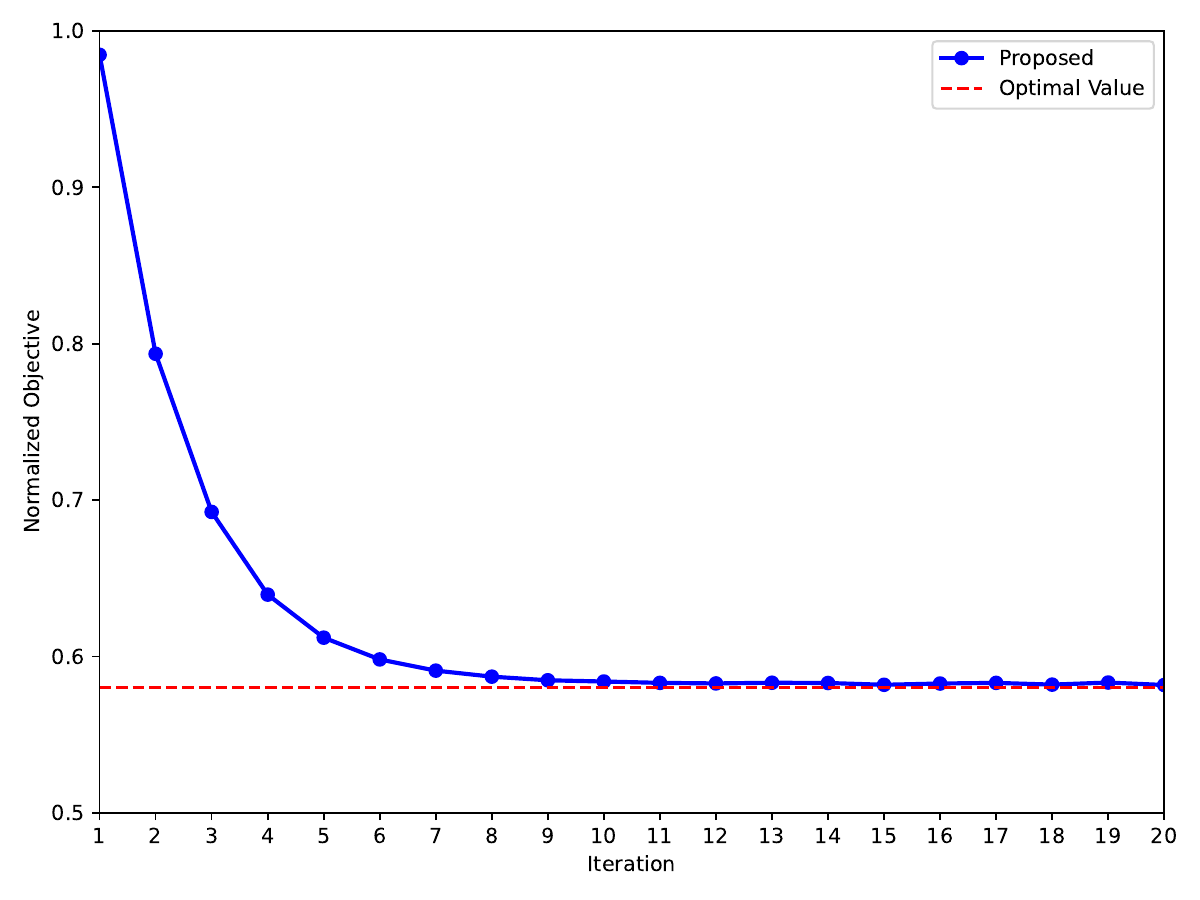}} 
\caption{Convergence of the proposed algorithm.} 
\label{f1} \end{figure}

Fig. \ref{f1} shows the convergence behavior of the proposed algorithm. We consider a normalized version of the objective function in \eqref{eqn13:main}. As you can observe from the figure, the proposed algorithm is converging smoothly and reaching a point near to the optimal in just 10 iterations.

\begin{figure}[!ht] 
\centerline{\includegraphics[width=3.5 in]{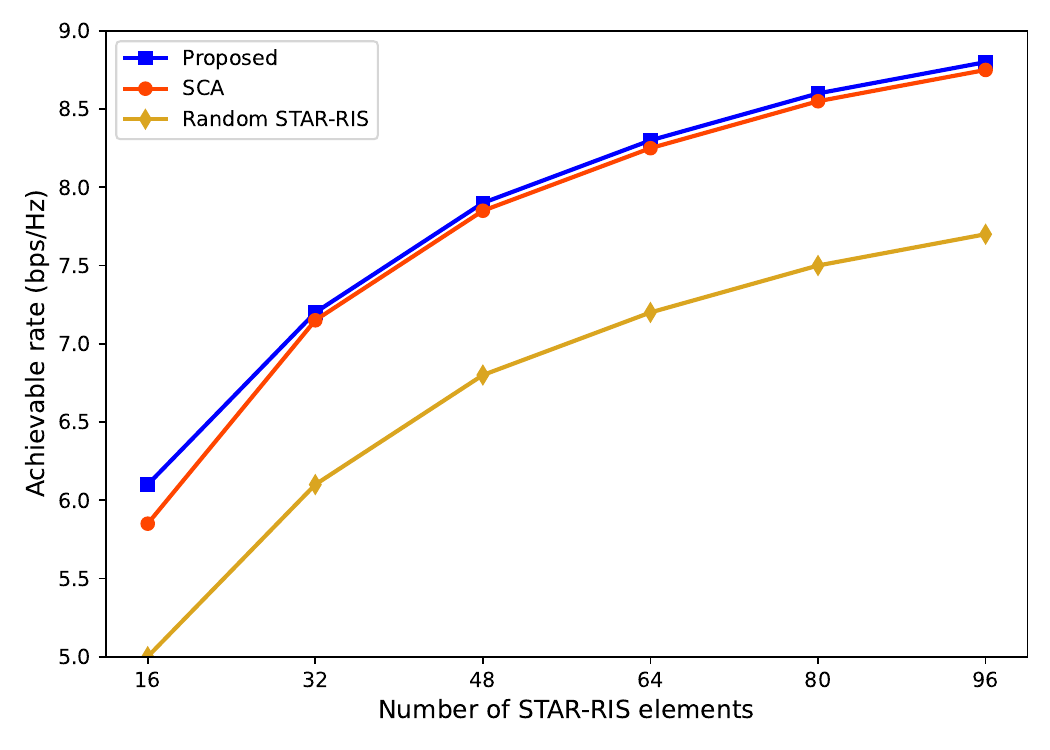}} 
\caption{Achievable rate versus the number of the STAR-RIS elements.} 
\label{f2} \end{figure}

Fig. \ref{f2} presents a comparison of the achievable rate as a function of the number of STAR-RIS elements. The performance advantage of the proposed algorithm over the SCA highlights the accuracy of the proposed algorithm. Furthermore, the comparison with the random STAR-RIS baseline demonstrates the effectiveness of jointly optimizing the transmission and reflection phase shifts. It is observed that the uplink data rate improves across all schemes as the number of passive elements increases, confirming the value of deploying more elements at the STAR-RIS until a saturation point is reached. Overall, the superior spectral efficiency achieved by the proposed method validates the importance of element-wise optimization of the phase configuration in STAR-RIS-assisted networks.

\section{Conclusion}
This paper has addressed the problem of real-time interference mitigation in STAR-RIS-aided multiuser wireless networks under practical constraints of statistical CSI and phase noise. We considered a realistic scenario where the STAR-RIS phase shifts are corrupted by random hardware-induced errors, modeled using the Von Mises distribution. To address the resulting nonconvex optimization challenge, we derived a closed-form approximation of the effective channel and reformulated the interference minimization problem on the complex circle Riemannian manifold.
We developed a Riemannian conjugate gradient algorithm tailored to the geometry of the problem, enabling efficient and unconstrained updates of the STAR-RIS configuration. The proposed method not only accommodates phase noise and statistical CSI but also achieves real-time optimization performance suitable for dense multiuser environments. Simulation results demonstrated the effectiveness of our approach in minimizing inter-user interference and improving SINR, with significant advantages in convergence and robustness compared to traditional techniques such as SCA.
Future work may extend this framework to hybrid RIS architectures, joint transmit-RIS beamforming, and mobility-aware optimization under dynamic user conditions.

\section*{Appendix I}
\section*{Proof of Theorem 1}
To derive the closed-form expression for $\mathbb{E}\left\{ {{{\bf{G}}_{{\rm eff},k,j}}} \right\}$, we find the approximation for $\mathbb{E}\left\{ {{\bf{h}}_{bk}^H} \right\}$ and  $\mathbb{E}\left\{ {{\bf{H}}_{br}^{i,H}{\Theta ^i}{\Phi ^i}{\bf{h}}_{rk}^i} \right\}$. Let ${\varpi _{mq}} = \exp \left( {j\frac{{2\pi }}{\lambda }a\left( {m - 1} \right)\left( {\sin \phi _{bk}^q} \right)} \right)$ and  ${\varsigma _{mq}} = {\Re _{mq}} + j{\Im _{mq}}$, then using \cite[Lemma~1]{6816003}, we can rewrite $\mathbb{E}\left\{ {{\bf{h}}_{bk}^H} \right\}$ as follows
\begin{equation}\label{Aeqn1}
\mathbb{E}\left\{ {{\rm{h}}_{bk}^H} \right\} = \frac{{\sqrt {{\beta _{bk}}} {\Lambda _m} + {\xi _m}}}{{\sqrt {{\beta _{bk}} + 1} }},
\end{equation}
where
\begin{equation*}
\Lambda _m^k = \sum\limits_{m = 1}^M {{\varpi _{mq}}}  = \sum\limits_{m = 1}^M {\cos \left( {\frac{{2\pi }}{\lambda }a\left( {m - 1} \right)\sin \phi _{bk}^m} \right)} ,
\end{equation*}
\begin{equation*}
\xi _m^k = \sum\limits_{m = 1}^M {{\varsigma _{mq}}} ,
\end{equation*}
Similarly, we can derive the approximate expression for $\mathbb{E}\left\{ {{\bf{H}}_{br}^{i,H}{\Theta ^i}{\Phi ^i}{\bf{h}}_{rk}^i} \right\}$ using \cite[Lemma~3]{6816003} as follows
\begin{equation}\label{Aeqn2}
 \mathbb{E}\left\{ {{\bf{H}}_{br}^{i,H}{\Theta ^i}{\Phi ^i}{\bf{h}}_{rk}^i} \right\}= \frac{{\sqrt {{\beta _{br}}} \sqrt {{\beta _{rk}}} \Lambda _{mn}^i + \sqrt {{\beta _{br}}}  + \sqrt {{\beta _{rk}}}  + 1}}{{\sqrt {\left( {{\beta _{br}} + 1} \right)\left( {{\beta _{rk}} + 1} \right)} }},
\end{equation}
where
\begin{equation*}
\resizebox{0.97\hsize}{!}{$\Lambda _{mn}^{i,k} = MN - 2\chi \frac{{\sum\limits_{1 \le q < n \le N}^N {\cos \left( {\theta _n^i - \theta _q^i + \left( {\frac{{2\pi }}{\lambda }d\left( {\left( {{x_n} - {x_q}} \right){{\cal A}_{rk}} + \left( {{z_n} - {z_q}} \right){{\cal B}_{rk}}} \right)} \right)} \right)} }}{{\sum\limits_{mn = 1}^{MN} {\cos \left( {\theta _n^i - \theta _m^i + \left( {\frac{{2\pi }}{\lambda }a\left( {\left( {{x_n} - {x_m}} \right){{\cal A}_{br}} + \left( {{z_n} - {z_m}} \right){{\cal B}_{br}}} \right)} \right)} \right)} }},$}
\end{equation*}
${{\cal A}_{rk}} = \sin {\phi _{rk}}\sin {\psi _{rk}}$, ${{\cal B}_{rk}} = \cos {\phi _{rk}}\sin {\psi _{rk}}$, ${{\cal A}_{br}} = \sin {\phi _{br}}\sin {\psi _{bk}}$, and ${{\cal B}_{br}} = \cos {\phi _{br}}\sin {\psi _{br}}$

End of proof.

\linespread{1.1}
\bibliographystyle{IEEEtran}
\bibliography{shpp}

\end{document}